\documentclass[12pt]{article}
\usepackage{amsmath}
\usepackage{graphicx,psfrag,epsf}
\usepackage{enumerate}
\usepackage{arydshln}
\usepackage{natbib}
\usepackage{url} 

\usepackage{lineno}
\usepackage{color}
\usepackage{ulem}

\newcommand{\blind}{0}

\begin{document}



\newcommand{\bA}{\ensuremath{\mathbf A}}
\newcommand{\bB}{\ensuremath{\mathbf B}}
\newcommand{\bE}{\ensuremath{\mathbf E}}
\newcommand{\bG}{\ensuremath{\mathbf G}}
\newcommand{\bX}{\ensuremath{\mathbf X}}
\newcommand{\bU}{\ensuremath{\mathbf U}}
\newcommand{\bV}{\ensuremath{\mathbf V}}
\newcommand{\bD}{\ensuremath{\mathbf D}}
\newcommand{\bR}{\ensuremath{\mathbf R}}
\newcommand{\bC}{\ensuremath{\mathbf C}}
\newcommand{\bH}{\ensuremath{\mathbf H}}
\newcommand{\bI}{\ensuremath{\mathbf I}}
\newcommand{\bJ}{\ensuremath{\mathbf J}}
\newcommand{\bW}{\ensuremath{\mathbf W}}
\newcommand{\bZ}{\ensuremath{\mathbf Z}}
\newcommand{\bL}{\ensuremath{\mathbf L}}
\newcommand{\bO}{\ensuremath{\mathbf O}}

\newcommand{\ba}{\ensuremath{\mathbf a}}
\newcommand{\bbb}{\ensuremath{\mathbf b}}
\newcommand{\be}{\ensuremath{\mathbf e}}
\newcommand{\bff}{\ensuremath{\mathbf f}}
\newcommand{\bg}{\ensuremath{\mathbf g}}
\newcommand{\bp}{\ensuremath{\mathbf p}}
\newcommand{\bq}{\ensuremath{\mathbf q}}
\newcommand{\bx}{\ensuremath{\mathbf x}}
\newcommand{\bhat}[1]{\mathbf{\hat{\rm #1}}}
\newcommand{\bteta}{\ensuremath{\boldsymbol \theta}}
\newcommand{\bPsi}{\ensuremath{\mathbf \Psi}}

\newcommand{\citeyearp}[1]{(\citeyear{#1})}
\newcommand{\norm}[1]{\parallel\! #1 \!\parallel}
\newcommand{\pref}[1]{(\ref{#1})}
\newcommand{\trace}[1]{\mbox{tr}(#1)}
\newcommand{\bone}{\ensuremath{\mathbf 1}}
\newcommand{\bzero}{\ensuremath{\mathbf 0}}
\newcommand{\half}{\ensuremath{\frac{1}{2}}}

\newcommand{\tred}[1]{\textcolor{red}{#1}}
\newcommand{\tgre}[1]{\textcolor{green}{#1}}
\newcommand{\tblue}[1]{\textcolor{blue}{#1}}

\def\spacingset#1{\renewcommand{\baselinestretch}%
{#1}\small\normalsize} \spacingset{1}


  \title{\bf On the visualisation of the correlation matrix}
  \author{Jan Graffelman \hspace{.2cm}\\
    Department of Statistics and Operations Research,\\
    Universitat Polit\`ecnica de Catalunya\\
    Department of Biostatistics,\\
    University of Washington}
  \maketitle

\bigskip
\begin{abstract}
Extensions of earlier algorithms and enhanced visualization techniques for approximating a correlation matrix are presented. 
The visualization problems that result from using column or colum--and--row adjusted correlation matrices, which give
numerically a better fit, are addressed. For visualization of a correlation matrix a weighted alternating least
squares algorithm is used, with either a single scalar adjustment, or a column-only adjustment with symmetric factorization; these
choices form a compromise between the numerical accuracy of the approximation and the comprehensibility of the obtained
correlation biplots. Some illustrative examples are discussed.
\end{abstract}

\noindent%

\vfill

\newpage
\spacingset{1.45} 
\section{Introduction}
\label{sec:01}

\begin{figure}[htb]
  \centering
  \includegraphics[width=.80\textwidth]{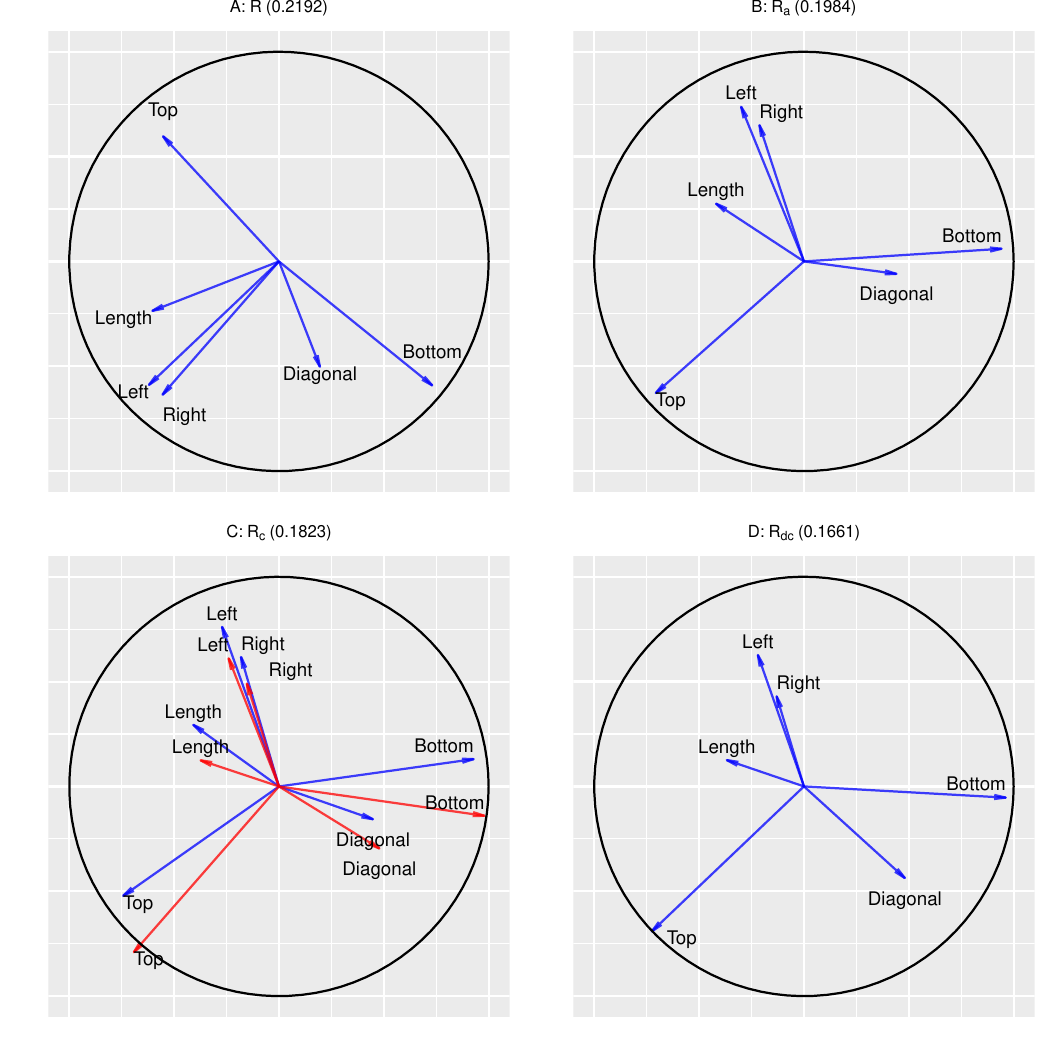}
  \caption{Visualizations of the correlation matrix obtained by standard \textsc{pca} (panel A), overall centering (panel B), column centering (panel C) and double centering (panel D).}
\label{fig:01}
\end{figure}

The correlation matrix is of fundamental importance in many scientific studies that use multiple variables. The visualization of correlation structure is therefore of great interest. In a recent article~\citep{Graffel40}, multivariate statistical methods for the visualization of the correlation matrix were reviewed, and a low-rank approximation with a scalar adjustment,
obtained by a weighted alternating least squares algorithm was proposed. This was shown to improve the approximation of the correlation matrix in comparison with existing methods like principal component analysis (\textsc{pca}) in terms of the root mean squared error (\textsc{rmse}). The previous work~\citep{Graffel40} capitalized on simplicity and symmetry of the visualization. In this article, it is shown that approximations of the correlation matrix with lower \textsc{rmse} are possible, but that these come at the price of sacrificing either simplicity or symmetry or both. One thus has to seek a balance between the {\it numerical exactness} of the approximation on one hand, and the {\it comprehensibility of the visualization} on the other. Simplicity refers to using one single vector (or point) to represent each variable. Doubling the number of points per variable quite obviously gives more flexibility in approximating the correlation matrix, but comes at the price of more dense and less attractive visualizations. Symmetry refers to symmetry of the approximation, in the sense that projecting vector A onto B leads to the same approximation to the correlation as projecting B onto A. To make these principles clear, four biplots of the correlation matrix of the banknote data are shown in
Figure~\ref{fig:01}; this data set consists of six size measurements ({\it Top, Bottom, Length, Left, Right} and {\it Diagonal}) on a sample of Swiss banknotes (\cite{Weisberg}, using counterfeits only). At this point, we do not worry about avoiding the fit of the diagonal of the correlation matrix, something that will be efficiently dealt with by the weigthed alternating least squares (\textsc{wals}) algorithm later. Figure 1A shows the "crude" approximation to $\bR$ obtained by \textsc{pca}. Its origin represents zero correlation for all variables. Figure 1B shows the biplot obtained after adjustment by a single scalar $\delta$, here taken to be the overall mean of the correlation matrix ($\delta = \bone' \bR \bone/p^2$), and represents $\bR_a = \bR - \delta \bJ$. Its origin represents correlation $r_{..}$ ($0.20$) for all variables, where we use the dot subindex to indicate averaging over the corresponding index. Figure 1C gives the biplot after single-centring or adjustment by column means ($\bR_c = \bH \bR$, with
idempotent centring matrix $\bH = \bI - (1/p) \bone \bone'$); its origin represents, for the $i$th variable, $r_{i.}$. Finally, Figure 1D shows the biplot after after double-centring or adjustment by both row and column means ($\bR_{dc} = \bH \bR \bH$) where the $i$th variable has origin $r_{i.} + r_{.j} - r_{..}$.  In each case, the biplot is obtained by the singular value decomposition (\textsc{svd};~\cite{Eckart}) of the corresponding adjusted correlation matrix.

\begin{figure}[htb]
  \centering
  \includegraphics[width=.80\textwidth]{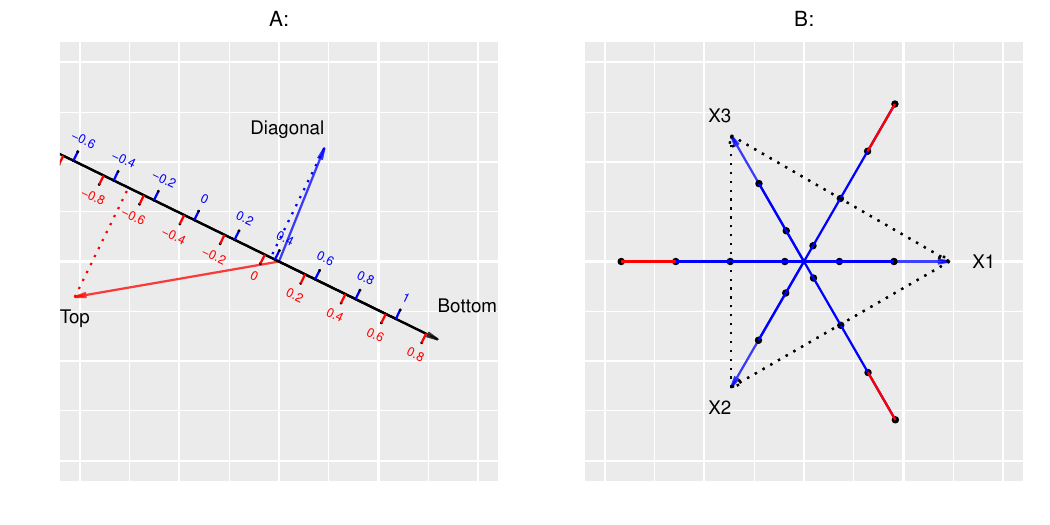}
  \caption{Interpretation of a biplot of the double-centered correlation matrix by the scalar product rule. A: An empirical $3 \times 3$
    correlation matrix with double calibration of the variable {\it Bottom}. The lower scale refers to the projection of {\it Top},
    the upper scale to the projection of {\it Diagonal}. B: An equicorrelation matrix with $r_{ij} = 0.20$. Black dots represent
    increments of 0.2 in the correlation scale.}
\label{fig:02}
\end{figure}

The \textsc{rmse} of the approximation is seen to decrease as we successively use $0, 1, p$ and $2p$ adjustment parameters. The \textsc{pca} biplot respects simplicity and symmetry, though it gives the worst approximation. Scalar adjustment retains simplicity and symmetry, and improves the approximation. Column adjustment destroys both simplicity and symmetry, but improves the approximation. In the corresponding biplot (Fig.~\ref{fig:01}C), correlations are approximated by scalar products between row coordinates (red) and column cooordinates (blue). Within-set scalar products (blue with blue or red with red) should not be considered. 
Row--and--column adjustment retains both
simplicity and symmetry, but complicates the interpretation of the scalar products.  This complication of the visualization is shown in more detail in Figure~\ref{fig:02} which shows a biplot of the double centred correlation matrix of three selected variables ({\it Diagonal, Top} and {\it Bottom}). Note that the double-centered correlation matrix has rank $p - 1$ and the three variable case is therefore perfectly represented in two-dimensional space (\textsc{rmse} = 0). However, for interpretation, a scalar product ($\bg_i' \bg_j$) in the plot needs to be backtransformed to the correlation scale, which implies the addition of the $i$th row mean, the $j$th column mean, and the subtraction of the overall mean of the original correlation matrix. The effect of the backtransformation can be shown by calibration of the biplot axis~\citep{Graffel17,Gower4,Gower6}. The backtransformation leads to a different calibration of the biplot vector for each scalar product. Figure~\ref{fig:02}A shows the double calibration of variable {\it Bottom}, where the lower scale (red) is for the projection of {\it Top}, and the upper scale (blue) for the projection of
{\it Diagonal}. The sample correlations of {\it Bottom} with these varables (-0.68 and 0.38 respectively) are perfectly recovered when referred to their corresponding scale, but result to huge errors when referred to the wrong scale. Note the different position of the zero correlation point on both scales. It is clear that the visualization of the correlation matrix by row {\it and} column adjustment (i.e., double centring) is highly complicated, because there is no common zero correlation point for the projections onto each biplot vector. An exception to these problems are those correlation matrices that have a (close to) constant margin, i.e., the same average correlation for each variable, such as an equicorrelation matrix. These can still be unambigously visualized despite the double-centring as is shown for a $3 \times 3$ equicorrelation matrix ($r_{ij} = 0.20, \forall i \not = j$) in Figure~\ref{fig:02}B, where the origin represents the overall correlation (0.47), and all between-variable projections are 0.20. This results from the equality of the overall mean and the marginal means of the equicorrelation matrix, such that the origin for the $i$th variable is $r_{i.} + r_{.j} - r_{..} = r_{i.}$ which does not depend on $j$, and is in fact identical for all variables.

The remainder of this article presents some theory on possible approximations to the correlation matrix based on weighted alternating least-squares algorithms, and provides software to generate the approximations and visualizations. Visualizations are integrated into the framework of the grammar of graphics~\citep{Wilkinson} making use of the R-package ggplot2~\citep{Wickham}.

\clearpage

\section{Materials and methods}
\label{sec:02}

We develop methods for the approximation of a correlation matrix by weighted alternating least squares algorithms, and design an
algorithm that approximates the correlation matrix while adjusting for column effects. Correlation tally sticks are presented to
facilitate the visualization of the approximations. Some generalized root mean squared error (\textsc{rmse}) statistics are proposed that
allow the calculation of overall and per-variable goodness-of-fit measures for both symmetric and asymmetric representations that use
weights. Several data sets are used to illustrate the approximations. 

\paragraph{Optimization problem}

The representations of the correlation matrix in the introduction are suboptimal, for they include the diagonal of ones. The fit of the diagonal can be avoided by using a weighted approach using $p \times p$ weight matrix $\bW$ with zeros on the diagonal. We seek a symmetric low rank approximation $\bR \approx \bG \bG'$ without fitting its diagonal, by minimizing the loss function

\begin{equation}
  \sigma(\bG) = \sum_{i=1}^{p} \sum_{j=1}^p w_{ij} (r_{ij} - {\bg_i}' \bg_j)^{2}.
  \label{eq:loss01}
\end{equation}

For equal weights ($\bW = \bJ = \bone \bone'$), Eq.~\pref{eq:loss01} is solved by the eigenvalue-eigenvector decomposition of $\bR$, which amounts
to a \textsc{pca} of the standardized data matrix. The scalar, column and column--plus--row adjustments presented in the Introduction
by using averages can be carried out more generally by minimizing the loss function

\begin{equation}
  \sigma(\bG,\delta,\bp,\bq) = \sum_{i=1}^{p} \sum_{j=1}^p w_{ij} (r_{ij} - \delta - \bone \bq' - \bp \bone' - {\bg_i}' \bg_j)^2,
   \label{eq:loss02}
\end{equation}

where $\delta$ represents a single scalar adjustment, $\bq$ a vector of column adjustments and $\bp$ a vector of row adjustments. The more general version of loss function~\pref{eq:loss02}, with the factorization $\bR \approx \bA \bB'$ is given by 

\begin{equation}
  \sigma(\bA,\bB,\delta,\bp,\bq) = \sum_{i=1}^{p} \sum_{j=1}^p w_{ij} (r_{ij} - \delta - \bone \bq' - \bp \bone' - {\ba_i}' \bbb_j)^2,
   \label{eq:loss03}
\end{equation}

and can be efficiently minimized with the R function {\tt wAddPCA}~\citep{DeLeeuw2022}. Program {\tt wAddPCA} allows for all possible
adjustments, as either $\delta$, $\bq$ or $\bp$ can be set to zero. To avoid the problems with the biplot interpretation when row
{\it and} column adjustment is performed (see Figure~\ref{fig:02}A) one can best either minimize these loss functions using either a
single scalar adjustment or a column adjustment only. The latter implies each variable has a single adjustment ($\delta + q_j$) which is represented by the origin of the
biplot. This is akin to an ordinary cases by variables \textsc{pca} biplot where the origin has the interpretation of representing the mean
of each variable, though it differs from \textsc{pca} because the diagonal is turned off by assigning it zero weight.
In terms of the calibration of a biplot axis, it means that each biplot vector has a unique zero correlation point, which favours
the visualization. Moreover, we suggest to require symmetry, such that $\bA = \bB$, meaning that we minimize Eq.~\pref{eq:loss02}
with $\bp = \bzero$ and seek the decomposition

\begin{equation}
\bR = \delta \bone \bone' + \bone \bq' + \bG \bG' + \bE.
\end{equation}

We note that this gives, counterintuitively,  an {\it asymmetric} approximation to $\bR$, but with a {\it symmetric} low rank factorization ($\bG \bG'$),
such that there is only a single biplot vector per variable. Note that the asymmetric nature of the approximation implies that
$\bE + \bone \bq'$ will be symmetric. Because there is an adjustment for each column, we expect that the fit
can be better than is obtained by a single scalar adjustment only (i.e.\ $\bq = \bzero$). The loss function, using the $p \times p$
non-negative unit-sum weight matrix $\bW$, can be concisely written as

\begin{equation}
\sigma(\bG,\delta,\bq) = \trace{\bE' (\bW \circ \bE)},
\end{equation}

where $\circ$ represents the Hadamard product, and $\bE$ a $p \times p$ matrix of residuals. Setting first order
derivatives w.r.t. $\delta$, $\bq$ and $\bC$ equal to zero, we respectively obtain 

\begin{equation}
\delta = \trace{\bW \circ \bR} -  \trace{\bq \bone' \bW}  - \trace{\bW \bG \bG'}, 
\label{eq:01}
\end{equation}

\begin{equation}
\bq = \bD_w^{-1} \left( \bW \circ ( \bR - \delta \bJ - \bG \bG') \bone \right),
\label{eq:02}
\end{equation}

where $\bD_w = diag(\bW \bone)$, and

\begin{equation}
(\bW \circ (\bR - \delta \bJ - \frac{1}{2}(\bone \bq') - \frac{1}{2}(\bq \bone') - \bG \bG')) \bG = \bO.
\label{eq:03}
\end{equation}

Importantly, we note the last equation can be resolved by applying  the {\tt ipSymLS} algorithm~\citep{DeLeeuw} to the adjusted form of the
correlation matrix

\begin{equation}
\bR_a = \bR - \delta \bJ - \frac{1}{2}(\bone \bq') - \frac{1}{2}(\bq \bone') - \bG \bG',
\end{equation}

for given $\delta$ and $\bq$.

\paragraph{Algorithm}

An iterative algorithm can now be set up by:

\begin{enumerate}
\item Choosing initial values for $\delta$ (e.g.\, $\delta = 0$) and $\bq$ (e.g.\, the column means of $\bR$).
\item Calculate $\bG$, resolving Eq. (\ref{eq:03}) with {\tt ipSymLS} for given rank of interest.
\item Update $\delta$ according to Eq. (\ref{eq:01}).
\item Return to step 2. and iterate until convergence.
\item Update $\bq$ according to Eq. (\ref{eq:02}).
\item Return to step 2. and iterate until convergence. 
\end{enumerate}

These steps have been programmed in function {\tt FitRDeltaQSym} of the R package Correlplot~\citep{GraffelCP}. A biplot of the adjusted
correlation matrix is obtained by plotting the rows of $\bG$. In the remainder, we will refer to the different options of the \textsc{wals}
algorithm as \textsc{wals}-null (no adjustments), \textsc{wals}-$\delta$ (only a scalar adjustment), \textsc{wals}-$q$ (only column adjustment), \textsc{wals}-$q$-sym (only column adjustment but with symmetric factorization (Eq. 4)) and \textsc{wals}-$p$-$q$ (row and column adjustment). 

\paragraph{Zero correlation point}

For the purpose of interpretation, it is convenient to find the point of zero correlation on each biplot vector, and mark it on each
biplot vector. After convergence, we have the approximation

\begin{equation}
\bR - \delta \bJ - \bone \bq' \approx \bG \bG'
\end{equation}

which implies

\begin{equation}
  r_{ij} - \delta - q_j \approx \bg_i' \bg_j =  || \bg_i|| || \bg_j|| \cos{\theta} = || \bg_j|| || \bp_{ij}||,  
\end{equation}

where $\bp_{ij}$ is the projection of $\bg_i$ onto $\bg_j$. This projection is proportional to $\bg_j$, and given by

\begin{equation}
\bp_{ij} = \frac{\bg_i'\bg_j}{\bg_j' \bg_j} \bg_j \approx \frac{r_{ij} - \delta - q_j}{\bg_j' \bg_j} \bg_j.
\end{equation}

By substituting $r_{ij} = 0$ in the latter equation, the coordinates of the zero correlation point on biplot vector $\bg_j$ are
found. Obviously, the coordinates of other values of interest (1, -1, etc.) can be found in the same way. In matrix terms, all
zero points can be found simultaneously with the expression

\begin{equation}
- diag(\delta + \bq) \bD_g^{-1} \bG \quad \mbox{ with } \bD_g = diag(\bG \bG').
\end{equation}

The latter equations can be used to create an equally spaced {\it correlation tally stick}, where the values \sloppy{$(-1.0, -0.8, -0.6, \ldots, +1.0)$} are marked
off as small dots on the biplot vector to facilitate reading off the approximation of the correlations without cluttering up the
biplot with a
fully calibrated scale with all its tick marks and tick mark labels. Colouring the biplot vector according to the sign of the
approximated correlation (e.g., red for negative, blue for positive) further enhances the visualization (See
Figures~\ref{fig:02}B,~\ref{fig:03} and~\ref{fig:04}).

\paragraph{Goodness-of-fit}

We provide a generic formula for calculating the \textsc{rmse} of an approximation to the correlation matrix from the error matrix $\bE$,
which takes into account that the approximation may not be symmetric,

\begin{equation}
RMSE = \sqrt{\frac{\sum_{i = 1}^p \sum_{j = 1}^p w_{ij} e_{ij}^2}{\sum_{i=1}^p \sum_{j=1}^p w_{ij}}} = 
\sqrt{\frac{\sum_{i \not = j}^p  w_{ij} e_{ij}^2 + \sum_i w_{ii} e_{ii}^2}{\sum_{i=1}^p \sum_{j=1}^p w_{ij}}}.
\end{equation}

This expression allows for weighting by means of a symmetric non-negative weight matrix. The exclusion of the diagonal from the \textsc{rmse}
calculation is possible by setting the diagonal of the weight matrix to zero. Likewise, the per-variable \textsc{rmse}, $z_i$, can be calculated with 
\begin{equation}
z_i = \sqrt{\frac{\sum_{j=1}^p w_{ij} e_{ij}^2 + \sum_{j=1}^p w_{ji} e_{ji}^2 - w_{ii} e_{ii}^2}{ \sum_{j = 1}^p w_{ij} + \sum_{j = 1}^p w_{ji} - w_{ii}}} = \sqrt{\frac{\sum_{j=1}^p w_{ij} e_{ij}^2 + \sum_{j=1}^p w_{ji} e_{ji}^2 - w_{ii} e_{ii}^2}{ 2 \sum_{j = 1}^p w_{ij}  - w_{ii}}}.
\end{equation}

where the terms $w_{ii} e_{ii}^2$ and $w_{ii}$ occur to avoid double-counting the errors on the diagonal. The overall \textsc{rmse} obviously
relates to the per-variable \textsc{rmse}, and this relation is found to be

\begin{equation}
RMSE = \sqrt{ \frac{\frac{1}{2} \sum_i z_i^2 \left( 2 \sum_{j = 1}^p w_{ij}  - w_{ii} \right) + \frac{1}{2} \sum_i w_{ii} (r_{ii} - \hat{r}_{ii})^2}{\sum_{i = 1}^p \sum_{j = 1}^p w_{ij}}}.
\end{equation}

Whenever the diagonal is turned off, this simplifies to

\begin{equation}
RMSE = \sqrt{ \frac{\sum_i z_i^2 \left(\sum_{j = 1}^p w_{ij}\right)}{\sum_{i = 1}^p \sum_{j = 1}^p w_{ij}}} = \sqrt{ \frac{\sum_i z_i^2 w_{i.} }{\sum_{i = 1}^p w_{i.}}},
\end{equation}

where $w_{i.}$ is the total marginal weight for the $i$th variable. The latter expression is the square root of a weighted average of the squared \textsc{rmse}s of each variable.

\section{Results}
\label{sec:03}

We give a few examples of the method we have developed. Table~\ref{tab:01} shows the \textsc{rmse} statistics for a few correlation matrices, analysed in more detail below, obtained by ten different methods: \textsc{pca}, the \textsc{svd} of $\bR_o$ (adjustment by subtracting its overall mean), the \textsc{svd} of $\bR_c$ (adjustment by column-centring), the \textsc{svd} of $\bR_{dc}$ (adjustment by double-centring), principal factor analysis (\textsc{pfa}), and \textsc{wals} with different adjustments. All applications of \textsc{wals} used zero weights for the diagonal, and we consecutively used no adjustment (\textsc{wals}-null), scalar adjustment only (\textsc{wals}-$\delta$), column adjustment with symmetric factorization (\textsc{wals}-$q$-sym), column adjustment without symmetry
(\textsc{wals}-$q$) and adjustment of both rows and columns (\textsc{wals}-$p$-$q$). \textsc{pca} and the three \textsc{svd}s try to approximate the diagonal of the correlation matrix, and the corresponding \textsc{rmse} calculations include the diagonal, whereas for \textsc{pfa} and \textsc{wals} the diagonal was excluded from \textsc{rmse} calculations. Table~\ref{tab:01} list the methods in order of expected decrease of the \textsc{rmse}. For \textsc{wals}, non-compliance of that order is typically indicative of non-convergence of one the algorithms.

\begin{table}[ht]
\centering
\begin{tabular}{lrrr}
  \hline
       & \multicolumn{3}{c}{Dataset}\\ \cline{2-4}
Method & Goblets & Milks & Beans \\ 
  \hline
  \textsc{pca}     $\bR$       & 0.0696 & 0.1183 & 0.1761 \\ 
  \textsc{svd}     $\bR_o$     & 0.0749 & 0.0813 & 0.1950 \\ 
  \textsc{svd}     $\bR_c$     & 0.0440 & 0.0550 & 0.1202 \\ 
  \textsc{svd}     $\bR_{dc}$   & 0.0210 & 0.0431 & 0.0997\\ \hdashline  
  \textsc{pfa}                 & 0.0417 & 0.0515 & 0.1097 \\ 
  \textsc{wals}-nul.            & 0.0417 & 0.0514 & 0.1097 \\ 
  \textsc{wals}-$\delta$          & 0.0417 & 0.0497 & 0.1062 \\ 
  \textsc{wals}-$q$-sym        & 0.0186 & 0.0146 & 0.1034 \\ 
  \textsc{wals}-$q$            & 0.0197 & 0.0140 & 0.0991 \\ 
  \textsc{wals}-$p$-$q$            & 0.0018 & 0.0003 & 0.0693 \\ 
  \hline
\end{tabular}
\caption{Goodness-of-fit statistics (\textsc{rmse}) for three correlation matrices obtained by using different methods. \textsc{rmse}s for the \textsc{svd}s
  (above the dashed line) include the diagonal of the correlation matrix, for \textsc{pfa} and \textsc{wals} variants the diagonal was excluded.}
\label{tab:01}
\end{table}

\subsection{Archeological goblets data}

We analyse the correlation matrix of six size measurements (rim diameter {\it RD}; bowl width {\it BW}; bowl height {\it BH};
foot diameter {\it FD}; width at the top of the stem {\it SW}; and stem height {\it SH}) of a sample of 25 archeological goblets from
Thailand~\citep{Manly} and given in Table~\ref{tab:02}.

\begin{table}[ht]
\centering
\begin{tabular}{lcccc:cc}
  \hline
 & SH & FD & BW & BH & RD & SW \\ 
  \hline
SH & 1.000 & 0.910 & 0.797 & 0.858 & 0.588 & 0.289 \\ 
  FD & 0.910 & 1.000 & 0.829 & 0.843 & 0.675 & 0.487 \\ 
  BW & 0.797 & 0.829 & 1.000 & 0.839 & 0.623 & 0.581 \\ 
  BH & 0.858 & 0.843 & 0.839 & 1.000 & 0.346 & 0.251 \\ \hdashline
  RD & 0.588 & 0.675 & 0.623 & 0.346 & 1.000 & 0.690 \\ 
  SW & 0.289 & 0.487 & 0.581 & 0.251 & 0.690 & 1.000 \\ 
   \hline
\end{tabular}
\caption{Correlation matrix of Manly's Goblets data.}
\label{tab:02}
\end{table}

\begin{figure}[htb]
  \centering
  \includegraphics[width=\textwidth]{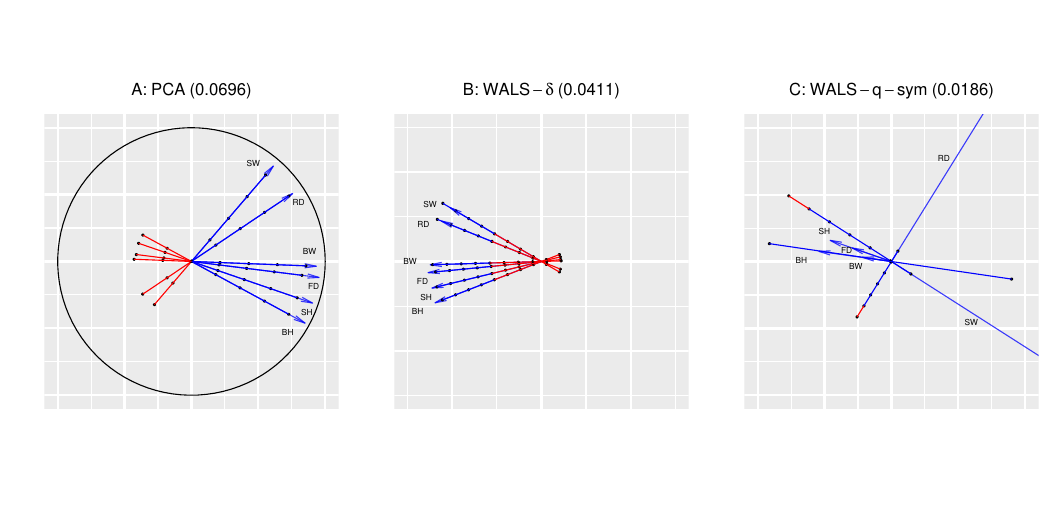}
  \caption{Biplots of archeological goblets. A: \textsc{pca} biplot of the correlation matrix. B: \textsc{wals}-$\delta$ biplot ($\delta = -0.72$).
    C: \textsc{wals}-$q$-sym biplot. Black dots represent increments of 0.2 in the correlation scale. Blue tails correspond to the positive part of the correlation scale, red tails to the negative part. The \textsc{rmse} of the approximation is given in the title of each panel.}
\label{fig:03}
\end{figure}

Figure~\ref{fig:03} shows three biplots of the correlation matrix obtained by a standard correlation-based \textsc{pca}~(\ref{fig:03}A),
\textsc{wals}-$\delta$~(\ref{fig:03}B) and \textsc{wals}-$q$-sym~(\ref{fig:03}C).
In this case, the last analysis gives the lowest \textsc{rmse} (0.0186). \textsc{pca} knots all biplot vectors at correlation zero
in the origin. \textsc{wals}-$\delta$ knots all biplot vectors at correlation $\delta = -0.72$ in the origin. In the last
analysis~(\ref{fig:03}C), each biplot vector has it own specific correlation represented by the origin, which gives additional flexibility
to fit the correlation matrix and allows for reduction of the \textsc{rmse} (see Table~\ref{tab:01}). Figure~\ref{fig:03}C displays a long stretching of the vectors
{\it RD} and {\it SW}, whose end points are outliers that are out of the plot. Consequently there is a high rate of change of
the correlation along these two vectors.
Standard correlation tally sticks with 0.2 unit changes are shown for {\it RD} and {\it SW}; an additional correlation stick for {\it BW}
is shown with black dots for correlations (0.75, 0.80 and 0.85). The rate of change of the correlation for the shorter vector {\it BW}, as
well as for the other three shorter vectors is much lower. Table~\ref{tab:03} shows all variables achieve smaller \textsc{rmse} in the last representation,
but that the stretching benefits {\it RD} and {\it SW} in particular, whose \textsc{rmse} is reduced about eight to ten-fold in comparison with
using a scalar adjustment only. The four short vectors for {\it SH}, {\it FD}
and {\it BW} with {\it BH} emphasize these variables form an approximately {\it equicorrelated block} in the correlation matrix, and that
their correlations with the stretched pair {\it RD} and {\it SW} show more variation. Correlations in Table~\ref{tab:02} were, a posteriori,
conveniently ordered to confirm this detected feature of the correlation matrix.

\begin{table}[ht]
 \scriptsize
\centering
\begin{tabular}{llrrrrrrrrrr}
  \hline
Dataset & Method & RD & BW & BH & FD & SW & SH \\ 
  \hline
Goblets & \textsc{pca}           & 0.0901 & 0.0637 & 0.0506 & 0.0384 & 0.0762 & 0.0535 \\ 
        & \textsc{wals}-$\delta$ & 0.0454 & 0.0427 & 0.0408 & 0.0278 & 0.0401 & 0.0471 \\ 
        & \textsc{wals}-$q$-sym  & 0.0044 & 0.0174 & 0.0110 & 0.0268 & 0.0051 & 0.0299 \\ 
   \hline
        &               & Density & Fat & Protein & Casein & Dry & Yield \\ 
  \hline
Milk    & \textsc{pca}           & 0.1692 & 0.0677 & 0.0912 & 0.0681 & 0.0831 & 0.1122 \\ 
        & \textsc{wals}-$\delta$ & 0.0683 & 0.0437 & 0.0238 & 0.0052 & 0.0621 & 0.0563 \\ 
        & \textsc{wals}-$q$-sym  & 0.0044 & 0.0147 & 0.0062 & 0.0134 & 0.0197 & 0.0209 \\
  \hline
        &               & Area & PM & MjAl & MiAL & AR & EXT & SOL & ROU & SF2 & SF4 \\ 
  \hline
Beans   & \textsc{pca}           & 0.0660 & 0.0798 & 0.0317 & 0.1596 & 0.2173 & 0.2118 & 0.1939 & 0.0865 & 0.1323 & 0.2110 \\ 
        & \textsc{wals}-$\delta$ & 0.0605 & 0.0653 & 0.0664 & 0.1444 & 0.1767 & 0.0444 & 0.1389 & 0.0523 & 0.1095 & 0.1116 \\ 
        & \textsc{wals}-$q$-sym  & 0.0503 & 0.0646 & 0.0636 & 0.1286 & 0.1528 & 0.0384 & 0.1523 & 0.0731 & 0.1159 & 0.1133 \\ 
   \hline
\end{tabular}
\caption{Per-variable goodness-of-fit statistic (\textsc{rmse}) for the datasets Goblets, Milk and Beans, using three methods.}
\label{tab:03}
\end{table}

\clearpage

\subsection{Milk data}

We analyse the correlation matrix of six variables registred on 85 milks ({\it Density, Fat, Protein, Casein, Dry} and {\it Yield}),
available in the FactoMineR package~\citep{Factominer} and based on a study by Daudin et al.~\citeyearp{Daudin}. The correlation matrix
of the variables is shown in Table~\ref{tab:04}.

\begin{table}[ht]
\centering
\begin{tabular}{lrrrrrr}
  \hline
 & Density & Fat & Protein & Casein & Dry & Yield \\ 
  \hline
Density & 1.000 & 0.402 & 0.543 & 0.596 & 0.753 & 0.432 \\ 
  Fat & 0.402 & 1.000 & 0.446 & 0.430 & 0.683 & 0.652 \\ 
  Protein & 0.543 & 0.446 & 1.000 & 0.958 & 0.668 & 0.641 \\ 
  Casein & 0.596 & 0.430 & 0.958 & 1.000 & 0.695 & 0.617 \\ 
  Dry & 0.753 & 0.683 & 0.668 & 0.695 & 1.000 & 0.700 \\ 
  Yield & 0.432 & 0.652 & 0.641 & 0.617 & 0.700 & 1.000 \\ 
   \hline
\end{tabular}
\caption{Correlation matrix of Daudin's milk data, as included in the FactoMineR package.}
\label{tab:04}
\end{table}

\begin{figure}[htb]
  \centering
  \includegraphics[width=\textwidth]{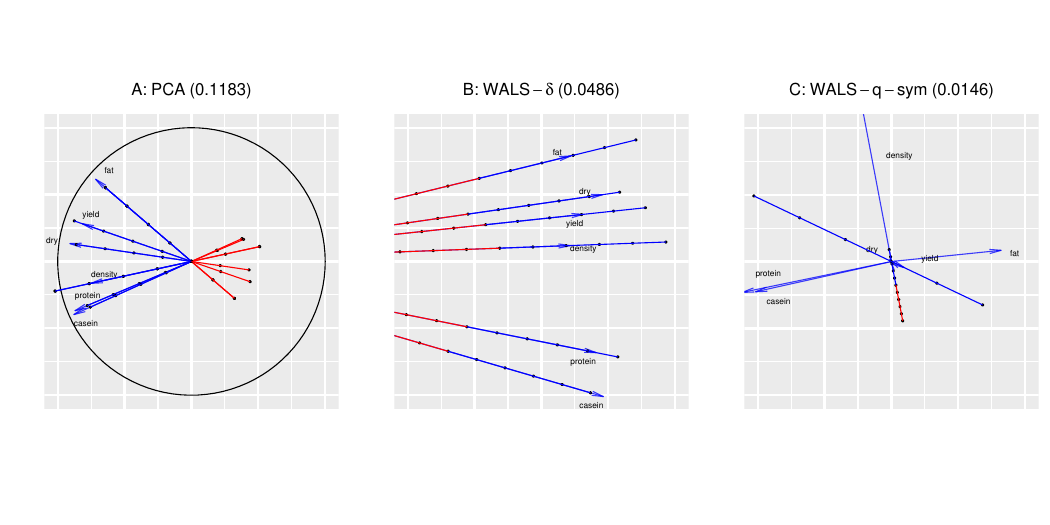}
  \caption{Biplots of milk data correlation structure. A: \textsc{pca} biplot of the correlation matrix. B: \textsc{wals}-$\delta$ biplot ($\delta = -2.16$). C: \textsc{wals}-$q$-sym biplot. Black dots represent increments of 0.2 in the correlation scale. Blue tails correspond to the positive part of the correlation scale, red tails to the negative part. The \textsc{rmse} of the approximation is given in the title of each panel.}
\label{fig:04}
\end{figure}

Figure~\ref{fig:04}
shows the three biplots obtained by \textsc{pca}, \textsc{wals}-$\delta$ and \textsc{wals}-$q$-sym. The tight relationship between {\it Protein} and {\it Casein} is visible
in all plots. The $\delta$ adjustment gives a considerable reduction in \textsc{rmse} with respect to \textsc{pca}, and the column-adjustment with symmetric
decomposition further reduces the \textsc{rmse}. Indeed, the latter two-dimensional representation of the correlation matrix is close to
perfect. In this case, when using the $\delta$ adjustment (Figure~\ref{fig:04}B), a large negative value of the adjustment parameter is obtained
($ \delta = -2.16$) which is out of the range of the correlation scale. This emphasizes that for this data, with considerable positive
correlations for all variables, the origin of the biplot is ultimately uninteresting; it is the zero in the correlation scale on the
biplot vector that is actually needed for adequate visualization and interpretation of the correlation structure. When a
column-adjustment with symmetric factorization is used, variable {\it Density} becomes an outlier (off the plot) and its biplot
vector is largely stretched and
consequently the projections onto this vector change fast as shown by the standard tally stick on {\it Density}. To the contrary,
variables {\it Dry} and {\it Yield} have very short biplot vectors such that their correlations hardly change across the biplot. To
illustrate this, Figure~\ref{fig:04}C shows the customized tally stick of {\it Yield} that spans the range (0.63, 0.68) with 0.01 increments.
This emphasizes, outlier {\it Density} taken apart, the approximate equicorrelation of {\it Dry} and {\it Yield} with all other variables.
The per-variable \textsc{rmse} in Table~\ref{tab:03} shows the variable that is most stretched, {\it Density}, experiments the largest reduction in \textsc{rmse} when the 
column-adjustment is used.

\clearpage

\subsection{Dry bean data}

We present an analysis of a larger correlation matrix of 16 measurements taken on 3546 dry beans of the variety Dermason~\citep{Koklu}
available at the UCI Machine Learning Repository (https://archive.ics.uci.edu/datasets). The 16 variables are Area, Perimeter ({\it PM}),
MajorAxisLength ({\it MjAL}), MinorAxisLength ({\it MiAL}), AspectRatio ({\it AR}), Eccentricity ({\it ECC}), ConvexArea ({\it CA}),
EquivDiameter ({\it ED}), Extent ({\it EXT}),
Solidity ({\it SOL}), Roundness ({\it ROU}), Compactness ({\it COM}) and
four Shape factors ({\it SF1, SF2, SF3} and {\it SF4}). These variables capture different aspects of the size and
the shape of the beans, and their correlation matrix is shown in Table~\ref{tab:05}.

\begin{table}[ht]
\tiny
\centering
\begin{tabular}{lrrrrrrrrrrrrrrrr}
  \hline
 & Area & PM & MjAL & MiAL & AR & ECC & CA & ED & EXT & SOL & ROU & COM & SF1 & SF2 & SF3 & SF4 \\ 
  \hline
Area & 1.00 & 0.97 & 0.92 & 0.91 & 0.13 & 0.15 & 1.00 & 1.00 & -0.01 & 0.23 & -0.04 & -0.14 & -0.90 & -0.67 & -0.14 & 0.04 \\ 
  PM & 0.97 & 1.00 & 0.95 & 0.83 & 0.25 & 0.26 & 0.97 & 0.97 & -0.06 & 0.05 & -0.27 & -0.26 & -0.83 & -0.74 & -0.26 & -0.05 \\ 
  MjAL & 0.92 & 0.95 & 1.00 & 0.68 & 0.50 & 0.51 & 0.92 & 0.92 & -0.08 & 0.16 & -0.25 & -0.50 & -0.68 & -0.90 & -0.50 & -0.00 \\ 
  MiAL & 0.91 & 0.83 & 0.68 & 1.00 & -0.29 & -0.27 & 0.91 & 0.91 & 0.07 & 0.25 & 0.19 & 0.28 & -1.00 & -0.31 & 0.28 & 0.06 \\ 
  AR & 0.13 & 0.25 & 0.50 & -0.29 & 1.00 & 0.98 & 0.13 & 0.13 & -0.19 & -0.08 & -0.57 & -1.00 & 0.29 & -0.81 & -0.99 & -0.08 \\ 
  ECC & 0.15 & 0.26 & 0.51 & -0.27 & 0.98 & 1.00 & 0.15 & 0.15 & -0.18 & -0.06 & -0.54 & -0.99 & 0.27 & -0.83 & -0.99 & -0.06 \\ 
  CA & 1.00 & 0.97 & 0.92 & 0.91 & 0.13 & 0.15 & 1.00 & 1.00 & -0.02 & 0.21 & -0.05 & -0.14 & -0.90 & -0.67 & -0.14 & 0.03 \\ 
  ED & 1.00 & 0.97 & 0.92 & 0.91 & 0.13 & 0.15 & 1.00 & 1.00 & -0.01 & 0.23 & -0.04 & -0.14 & -0.91 & -0.67 & -0.14 & 0.04 \\ 
  EXT & -0.01 & -0.06 & -0.08 & 0.07 & -0.19 & -0.18 & -0.02 & -0.01 & 1.00 & 0.14 & 0.21 & 0.19 & -0.07 & 0.14 & 0.19 & 0.09 \\ 
  SOL & 0.23 & 0.05 & 0.16 & 0.25 & -0.08 & -0.06 & 0.21 & 0.23 & 0.14 & 1.00 & 0.73 & 0.09 & -0.26 & -0.07 & 0.09 & 0.51 \\ 
  ROU & -0.04 & -0.27 & -0.25 & 0.19 & -0.57 & -0.54 & -0.05 & -0.04 & 0.21 & 0.73 & 1.00 & 0.57 & -0.21 & 0.44 & 0.57 & 0.38 \\ 
  COM & -0.14 & -0.26 & -0.50 & 0.28 & -1.00 & -0.99 & -0.14 & -0.14 & 0.19 & 0.09 & 0.57 & 1.00 & -0.29 & 0.82 & 1.00 & 0.11 \\ 
  SF1 & -0.90 & -0.83 & -0.68 & -1.00 & 0.29 & 0.27 & -0.90 & -0.91 & -0.07 & -0.26 & -0.21 & -0.29 & 1.00 & 0.31 & -0.28 & -0.09 \\ 
  SF2 & -0.67 & -0.74 & -0.90 & -0.31 & -0.81 & -0.83 & -0.67 & -0.67 & 0.14 & -0.07 & 0.44 & 0.82 & 0.31 & 1.00 & 0.82 & 0.05 \\ 
  SF3 & -0.14 & -0.26 & -0.50 & 0.28 & -0.99 & -0.99 & -0.14 & -0.14 & 0.19 & 0.09 & 0.57 & 1.00 & -0.28 & 0.82 & 1.00 & 0.10 \\ 
  SF4 & 0.04 & -0.05 & -0.00 & 0.06 & -0.08 & -0.06 & 0.03 & 0.04 & 0.09 & 0.51 & 0.38 & 0.11 & -0.09 & 0.05 & 0.10 & 1.00 \\ 
   \hline
\end{tabular}
\caption{Correlation matrix of the Dry bean data for variety Dermason.}
\label{tab:05}
\end{table}

\begin{figure}[htb]
  \centering
  \includegraphics[width=\textwidth]{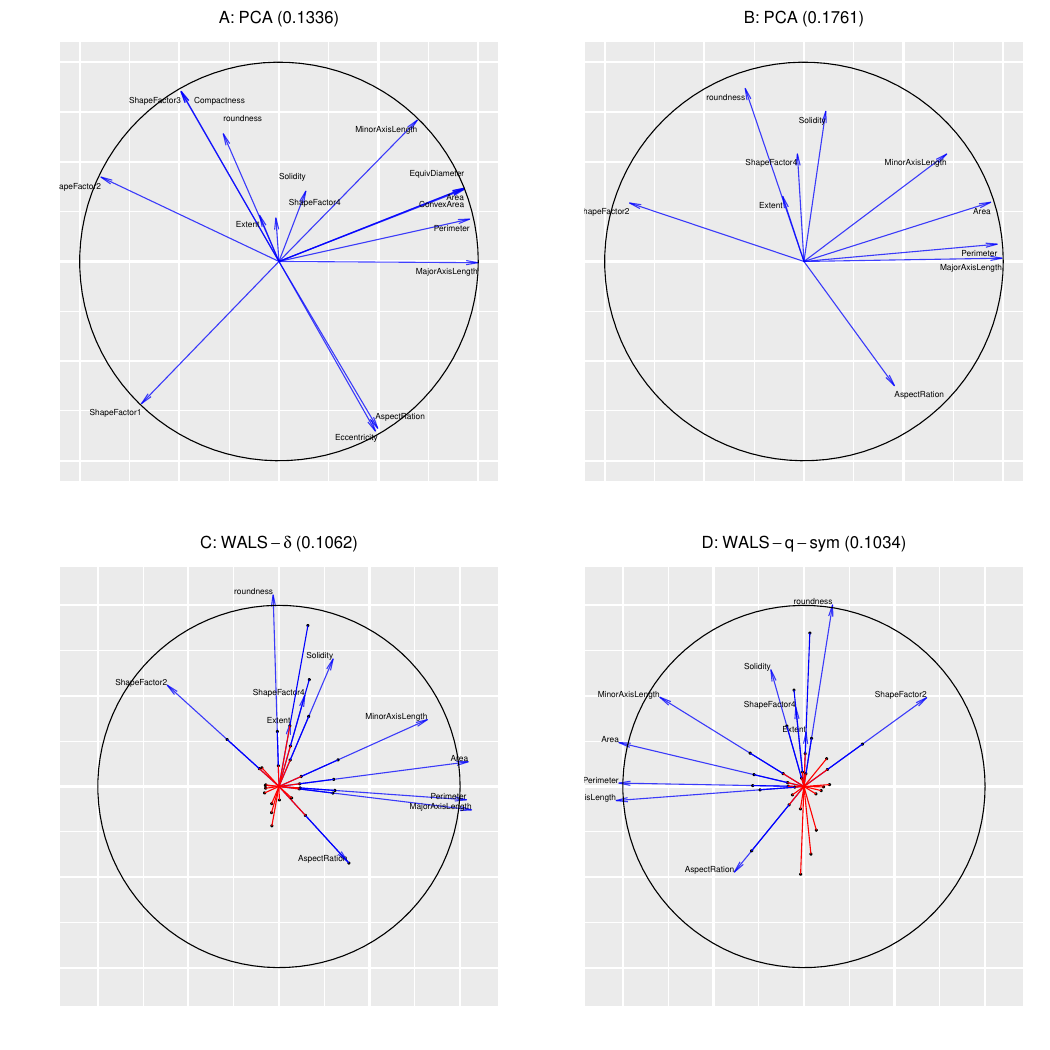}
  \caption{Biplots of the Beans data correlation structure. A: \textsc{pca} biplot of the original correlation matrix. B: \textsc{pca} biplot of the
    reduced correlation matrix. C: \textsc{wals}-$\delta$ biplot ($\delta = -0.12$). C: \textsc{wals}-$q$-sym biplot. Black dots represent increments of 0.2 in the correlation scale; only the (-0.2,0,0.2) part of the scale is indicated. Blue tails correspond to the positive part of the correlation scale, red tails to the negative part. The \textsc{rmse} of the approximation is given in the title of each panel.}
\label{fig:05}
\end{figure}

Biplots of the correlation structure are shown in Figure~\ref{fig:05}. Figure~\ref{fig:05}A shows a standard \textsc{pca} correlation biplot, which has a \textsc{rmse} of 0.1336. This reveals several variables are almost perfectly correlated, suggesting redundancy in some of the defined variables, where ({\it SF3, COM, ECC, AR}) apparently measure the same thing, as well as ({\it SF1, MiAl}), and ({\it Area, ED, CA}). We repeated \textsc{pca} retaining only one variable of each of these three groups, so reducing the number of variables to ten. Figure~\ref{fig:05}B shows the biplot with the reduced dataset, where the \textsc{rmse} has increased to 0.1761 due to the elimination of large part of the redundancy in the data. Figure~\ref{fig:05}C shows the biplot obtained by using the \textsc{wals}-$\delta$; this considerably improves the goodness-of-fit of the correlation matrix (\textsc{rmse} 0.1062); the origin of this plots represents $\delta = -0.12$. In general, correlations fitted by \textsc{wals}-$\delta$ are lower (i.e.\,\,closer to zero) than those obtained by \textsc{pca}.

The \textsc{wals}-$q$-sym biplot only slightly decreases the \textsc{rmse} (0.1034) in comparison with the \textsc{wals}-$\delta$ biplot. In this case, the column adjustment does not lead to a significant improvement of the approximation (See Table~\ref{tab:02}). The \textsc{wals}-$q$-sym biplot appears similar to the \textsc{wals}-$\delta$ when the latter is reflected in the vertical axis. However, the latter no longer has a common correlation for the origin, and some of the variables have slightly worsened their representation while others have slightly improved
(see detailed \textsc{rmse} statistics in Table~~\ref{tab:03}).

\clearpage

\section{Discussion}
\label{sec:04}

In this article we have considered approximations and visualizations of a fundamental matrix in multivariate analysis, the correlation matrix. From a purely numerical point of view, the best approximation, in the weighted least squares sense, is obtained by performing weighted alternating least squares with row and column adjustment (as illustrated by the examples in Table~\ref{tab:01}). This is natural, because the use of both row and column adjustment offers most flexibility to fit the correlations. The optimal approximation is however, particularly hard to visualize by using scalar products as has been illustrated in Figure~\ref{fig:02}A. We have previously reported~\citep{Graffel40} on the equivalence of the approximation of $\bR$ by \textsc{wals} and the analysis of the double-centered correlation matrix proposed by Hills~\citeyearp{Hills}, when unit weights are used. We note that, with the use of zero weights for the diagonal, the row and column adjustments obtained by \textsc{wals} are no longer equivalent to a double-centering the correlation matrix; in this case \textsc{wals} will give a lower \textsc{rmse} than is
obtained by Hills' double-centered analysis.\\

By using only a column adjustment but requiring a symmetric low-rank factorization, visualization of the correlations by means of scalar products in a biplot becomes feasible, as shown in the examples in Figures~\ref{fig:03}C,~\ref{fig:04}C and~\ref{fig:05}D. The examples show this representation tends identify equicorrelation structure, where variables that equicorrelate with others collapse to the origin. An inconvenience of this representation is that the approximation is asymmetric in the sense that the projection of $x_1$ onto $x_2$ gives a numerically different approximation to $r_{12}$ than the reverse projection. For the empirical examples we have studied so far, these differences tended to be small. The representation of the correlation matrix by a single scalar adjustment, as previously advocated~\citep{Graffel40}, does not suffer from this assymetry but will have a larger \textsc{rmse}. Admittedly, the decrease in \textsc{rmse} obtained by using a column adjustment with symmetric factorization was found to be small in most cases hitherto studied, and the $\delta$-only representation may therefore be favoured for retaining the symmetry of approximation. \textsc{wals}-$q$-sym has been developed here for the fundamental sake of
optimality, in order to identify the best approximation to the correlation matrix that can still be visualized. 

The column centring of the correlation matrix is inconvenient, because it complicates the visualization by doubling the number of biplot vectors as shown in Figure 1C. One may try to match the two sets of biplot vectors by Procrustes rotation~\citep{Gower3} or by the more flexible distance-based matching approach~\citep{DeLeeuw2011}. In the case of a perfect match it may be argued one set of biplot vectors could be omitted. In practice, such perfect matches were not observed.   

Column and/or row adjustments are common operations for many multivariate methods, where the data matrix is often centered (e.g., in \textsc{pca}) or sometimes double centred prior to the use of the spectral decomposition or the singular value decomposition. In these cases the optimal values for the adjustment parameters for columns or rows are known prior to analysis, in particular if no weighting is applied. For the analysis with a single scalar adjustment $\delta$, there is no closed form solution available for finding $\delta$. Estimating $\delta$ with the overall mean of the correlation matrix seems intuitive, but is suboptimal, and can even worsen the representation in comparison with standard \textsc{pca}, as the examples in Table~\ref{tab:01} clearly show. One needs to run the \textsc{wals} algorithm to obtain the optimal value for $\delta$.  

The use of correlation tally sticks in biplots of correlation structure is advocated, not only for biplots obtained by \textsc{wals} but also for the standard correlation biplots obtained by \textsc{pca}. They increase the information content of the plot without overcrowding it. Moreover, the rate of change of a correlation along a biplot vector can differ considerably across biplot vectors, and the tally stick makes
this clearly visible. 

\clearpage

\if0\blind
{
  \section{Acknowledgements}
  This work was supported by the Spanish Ministry of Science and Innovation and the European Regional Development Fund under grant PID2021-125380OB-I00 (MCIN/AEI/FEDER); and the National Institutes of Health under Grant GM075091.\\
} \fi

\if1\blind
{
\section{Acknowledgements}
} \fi

The author reports there are no competing interests to declare.



\bigskip
\begin{center}
{\large\bf SUPPLEMENTARY MATERIAL}
\end{center}

\begin{description}

\item[R-package Correlplot:] R-package {\tt Correlplot} (version 1.1.0) has been updated, and contains code to calculate the approximations to the correlation matrix and to create the graphics shown in the article, for both R's base graphics environment as well as for the
  ggplot graphics. R-package {\tt Correlplot} has a vignette containing a detailed example showing how to generate all graphical representations of the correlation matrix (GNU zipped tar file).

\end{description}

\bibliographystyle{agsm}
\bibliography{CorrelationArXivpreprint}

\end{document}